%% file: main.tex
\documentclass{article}

\usepackage[final]{neurips_2024}

\makeatletter
\renewcommand{\@notice}{}
\makeatother

\usepackage[T1]{fontenc}
\usepackage[utf8]{inputenc}
\usepackage{booktabs}
\usepackage{longtable}
\usepackage{algorithm}
\usepackage{algpseudocode}
\usepackage{amsmath,amssymb}
\usepackage{mathtools}
\usepackage{subcaption}
\usepackage{xspace}
\usepackage{multirow}
\usepackage[table]{xcolor}
\usepackage{hyperref}
\usepackage{natbib}

\newcommand{\sysname}{\textsc{ProxyDrift}\xspace}
\newcommand{\dimname}[1]{\textsf{#1}}
\newcommand{\dimval}[1]{{\small\textsc{#1}}}

\newcommand{\romark}[1]{\rotatebox{55}{\scriptsize #1}}
\definecolor{confblue}{RGB}{41,128,185}
\definecolor{catintent}{RGB}{31,78,182}
\definecolor{cattone}{RGB}{153,50,204}
\definecolor{catoutput}{RGB}{34,139,34}
\definecolor{catcontent}{RGB}{180,140,0}
\definecolor{catobs}{RGB}{100,100,100}
\definecolor{driftbad}{RGB}{200,40,40}
\definecolor{driftaverage}{RGB}{220,130,30}
\definecolor{driftgood}{RGB}{40,150,60}
\DeclareMathOperator{\MI}{MI}
\DeclareMathOperator{\JSD}{JSD}
\DeclareMathOperator{\AMI}{AMI}
\newcommand{\E}{\mathbb{E}}
\newcommand{\indep}{\perp\!\!\!\perp}

\begin{document}

\title{Privacy-Preserving Data Drift Detection and Recovery for Large-Scale LLM Applications via Proxy Representations}

\author{%
  Michael Levit \quad Josh Ledgard \quad Haoyu Dong \quad Vishwas Suryanarayanan \\
  Eyal Kolman \quad Sharon Tan \quad Qiang Gan \quad Vishal Chowdhary \\[0.4em]
  \normalfont Microsoft Corporation \\
  \normalfont\small\texttt{\{mlevit, ledgardjosh, hadong, visuryan,} \\
  \normalfont\small\texttt{eyalko, shartan, qigan, vishalc\}@microsoft.com}
}

\maketitle

\begin{abstract}
LLM applications deployed at scale face a
fundamental challenge: privacy constraints prevent direct inspection
of user interactions, making it difficult to obtain any
representative evaluation dataset or to track the
ongoing evolution of production traffic.  We present \sysname, a
framework that (i)~identifies and measures drift between production
traffic and offline evaluation sets, and (ii)~constructs and
refreshes those evaluation sets accordingly; all without access to
raw user data.  Our approach operates entirely on non-PII \emph{proxy
representations}: structured, multi-dimensional descriptors derived
from LLM-based classification of user interactions.  We introduce
(1)~a chance-calibrated, \emph{redundancy-aware} (RA)
alignment score that aggregates per-dimension drift
measurements via mutual information; (2)~a conditional sampler that generates synthetic proxies
respecting inter-dimensional dependencies; (3)~a roundtrip
consistency analysis that exposes generator/classifier disagreements
and guides proxy taxonomy refinement; and (4)~a feedback-linkage
analysis that ties per-dimension and per-value proxy distributions
to user satisfaction, surfacing actionable failure and success modes.
Serving hundreds of
millions of users, \sysname enables continuous drift monitoring and
targeted synthetic data generation without exposing sensitive user
data. Experiments confirm strong roundtrip consistency, discriminator-level indistinguishability of synthetic
queries from human queries, and tight end-to-end alignment ($\text{RA} \approx 0.9$) with
production.
\end{abstract}

\input{sections/introduction}
\input{sections/related}
\input{sections/proxy}
\input{sections/drift}
\input{sections/sampling}
\input{sections/deployment}
\input{sections/evaluation}
\input{sections/conclusion}

\section*{GenAI Usage Disclosure}
Large language models (GPT-5.X, Claude Opus~4.X) were used as
tools during the research: (1)~as the classification and generation
engines within the \sysname pipeline itself (the subject of this
paper); (2)~for light editing and grammar checking of the manuscript
text.  All experimental design, algorithmic contributions, analysis,
and interpretation are the work of the human authors.

\bibliographystyle{abbrvnat}
\bibliography{references}

\appendix
\input{sections/appendix}

\end{document}

%% file: sections/introduction.tex
\section{Introduction}\label{sec:intro}

LLM applications deployed in privacy-sensitive settings face a fundamental obstacle to offline
evaluation: strict compliance constraints prevent direct access to
user queries, system responses, and grounding documents.  Raw interaction
data cannot leave the production environment, so practitioners often
have no eyes-on view of real traffic at all.  This makes it difficult
to assemble a representative offline evaluation dataset in the first
place, and even harder to keep one aligned with traffic as user
populations shift, new workflows emerge, and interaction patterns
change~\cite{gama2014survey, lu2018learning}.  Without a way to
characterize production usage from telemetry alone, evaluation
datasets remain unverified static snapshots whose relevance to live
traffic is unknown and quietly degrades over time. 
This undetected misalignment directly distorts the
decisions that offline evaluation is meant to inform.  An offline set
that under- or over-represents a scenario systematically inflates or
deflates aggregate quality scores, hides real user pain points, and
tunes model and prompt iterations against a non-existent user.
Detecting and quantifying this drift is therefore a prerequisite for
trustworthy offline evaluation.

Traditional drift-detection methods that compare distributions over
raw inputs~\cite{gama2014survey, lu2018learning} are inapplicable in
this setting: the inputs themselves cannot be inspected.  We address
this with \sysname, a framework that enables both
(i)~privacy-preserving comparison of production and offline
distributions, and (ii)~the construction and refresh of offline
evaluation datasets that match production -- all without exposing raw
user data.  The key insight is to replace sensitive interaction data
with structured, multi-dimensional \emph{proxy representations}:
non-PII descriptors that capture the behavioral and linguistic
characteristics of user interactions (e.g., intent type,
communication style, domain, output expectations) without retaining
any verbatim content. Individual dimensions of these proxies are either observed directly (e.g., query length), 
or generated by an LLM classifier operating within the production compliance boundary and
can be safely persisted in telemetry.

Given proxy representations for both production traffic and offline
evaluation datasets, \sysname provides four integrated capabilities:
(1)~\textbf{drift identification and measurement} via a
chance-calibrated, redundancy-aware alignment score that
quantifies divergence per dimension and aggregates with
redundancy discounting;
(2)~\textbf{evaluation dataset construction} via a
Chow--Liu tree-based conditional sampler~\cite{chow1968} that
generates synthetic proxies preserving pairwise dimensional
dependencies, followed by LLM-based hydration into
natural-language queries;
(3)~\textbf{continuous monitoring} through an end-to-end pipeline
integrating telemetry collection, histogram aggregation, alignment
scoring, and dashboards; and
(4)~\textbf{taxonomy and quality diagnostics} from intrinsic
consistency analysis (driving schema refinement) and from linking
per-dimension and per-value distributions to user feedback.

\sysname has been deployed in a major cloud-based productivity suite
serving hundreds of millions of users, where it monitors multiple
application scenarios continuously and generates synthetic
evaluation data on a weekly cadence.  
The optimized taxonomy of dimensions and values exhibits high
drift alignment (RA $\approx 0.9$), high roundtrip consistency, and
generates synthetic queries that an unguided LLM discriminator cannot reliably
tell apart from human queries.


%% file: sections/related.tex
\section{Related Work}\label{sec:related}
\paragraph{Data Drift and Distribution Shift.}
The problem of distribution shift in deployed ML systems has been extensively
studied under various names: concept drift~\cite{gama2014survey, lu2018learning},
dataset shift~\cite{rabanser2019failing}, and label shift~\cite{lipton2018detecting}.
Most prior work assumes direct access to input features or model predictions for
statistical testing, a luxury unavailable under production privacy constraints. 
\sysname differs in detecting drift \emph{indirectly}, through aggregate comparisons
over structured proxy representations rather than raw data, and in using the same
substrate to construct production-aligned evaluation sets.

\paragraph{Synthetic Data Generation.}
Synthetic tabular data has been approached via GANs~\cite{xu2019modeling},
VAEs, and more recently LLMs~\cite{borisov2023language, jordon2022synthetic}.
Our conditional sampler targets discrete, multi-dimensional proxy objects with a
known schema, using Chow--Liu trees~\cite{chow1968} that are interpretable and
outperform neural generators on the cross-dimension dependency structure we care
about. Architecturally closest to our solution is Aug-PE~\cite{xie2024augpe}
which produces differentially private synthetic text via inference-time API access
alone; \sysname shares the principle of keeping private data inside the trust
boundary and pushing only structured aggregates out, but it uses LLM-classified
proxies to characterize the production distribution itself rather than to release
DP synthetic text for training.

\paragraph{LLM-Based Classification and Generation.}
Using LLMs for structured classification and controlled 
generation~\cite{ouyang2022training, zhao2026survey} is well-established; our
framework chains the two and uses the generation--classification cycle
(\S\ref{sec:eval-roundtrip}) as a novel intrinsic measure of how faithfully proxy
information is preserved end to end. Clio~\cite{tamkin2024clio} characterizes
millions of Claude conversations via AI-assistant-driven clustering with downstream
applications to economic-task usage~\cite{handa2025economic} and longitudinal
adoption~\cite{appel2025economicindex}; WildChat~\cite{zhao2024wildchat} releases
a 1M-conversation opt-in ChatGPT corpus; and TnT-LLM~\cite{wan2024tntllm} iteratively
induces a label taxonomy over chat-style conversations. \sysname fixes a 21-dimension
schema rather than inducing one, and consumes classifier outputs as a
chance-calibrated distributional fingerprint for drift measurement and synthetic-data
alignment rather than as qualitative insights or features for downstream supervised tasks.


%% file: sections/proxy.tex
\section{Proxy Representation and Classification}\label{sec:proxy}

The foundation of \sysname is a structured \emph{proxy representation}
that captures the behavioral properties of user interactions without
retaining verbatim content.

A proxy assigns each interaction a value (or set of values) along
$D$~categorical dimensions defined by a JSON schema; in our
deployment, $D = 21$ dimensions organized along the following axes:

\paragraph{Classified vs. Observable.} 
    Classified dimensions are obtained by running the interaction
    through an LLM classifier and require
    semantic understanding of the query text. For example, the user's
    intent, communication style, domain, or expected output format. 
    Observable properties are computed deterministically from
    interaction metadata without an LLM (e.g., word-count
    bucket, query language, and explicit
    grounding-file types). 
\paragraph{Nominal vs.\ Ordinal.}
\emph{Nominal} dimensions have unordered categorical values
(e.g., for output format: \dimval{Word}, \dimval{PowerPoint}, \dimval{PDF}).
\emph{Ordinal} dimensions have a meaningful value ordering
(e.g., for output length: \dimval{Brief} $<$ \dimval{Moderate} $<$ \dimval{Detailed}
$<$ \dimval{Extensive}).
The distinction matters for distance computation 
(cf. \S\ref{sec:drift}).
\paragraph{Single-valued vs.\ multi-valued.}
\emph{Single-valued} dimensions assign exactly one label per
interaction (e.g., for communication formality: one of
\dimval{Formal}, \dimval{Neutral}, or \dimval{Informal}).  \emph{Multi-valued} dimensions
assign a (confidence ordered) list of labels. Only nominal dimensions can be
multi-valued (e.g., the intent dimension might assign both
\dimval{Document Creation} and \dimval{Research} to a query that requests
both).

Every label carries a confidence score $s \in \{0, 1, \ldots, 5\}$,
where $s=0$ is reserved for the special value \dimval{Unknown} (classifier
unable to determine a label) and $s=5$ indicates maximum
confidence or deterministic observation.  Multi-valued dimensions
list their labels in decreasing confidence order.
Observable properties always carry maximum confidence. For presentation convenience,
we also group the dimensions into five semantic categories as indicated in Table~\ref{tab:schema-overview}.



The LLM-based classifier turns user queries into structured proxy objects.  It
operates within the production compliance boundary
via a dedicated augmentation workflow.  Given a user interaction
and a structured prompt defining the taxonomy (allowed values,
format constraints, scoring guidelines), the LLM produces a JSON
object conforming to the proxy schema.

A \emph{self-correction loop} ensures schema compliance: if the
LLM's output violates structural constraints (invalid values,
malformed tuples, missing dimensions), the specific violations are
described in natural language and fed back for correction.  This
loop runs up to $K=10$ attempts, progressively
resolving syntax errors without human intervention.

Additional extensions were introduced that generate proxies either
for the entire multi-turn interaction or for individual queries given
the context of all previous turns while also taking discourse break into account.


%% file: sections/drift.tex
\section{Drift Measurement}\label{sec:drift}

Given proxy distributions for production traffic and an offline
evaluation dataset, \sysname quantifies drift at three levels:
aggregate and per-dimension alignments, as well as actionable per-value diagnostics. 
Throughout,
$P = (p_1, \ldots, p_n)$ denotes the production (reference)
distribution and $O = (o_1, \ldots, o_n)$ denotes the evaluation
(observed) distribution over $n$ categories.


Among several candidate distance metrics (Total Variation, Wasserstein, Euclidean, \ldots), 
we selected Jensen--Shannon distance (JSD) as the default for categorical distributions 
because it reacts strongly to support mismatches. Indeed, a category heavily used in production but absent from
the evaluation set (or vice versa) incurs a large penalty. For reference distribution $P$ and observed distribution $O$ over $n$
categories, the normalized JSD is defined as
\begin{equation}\label{eq:jsd}
  \JSD(P, O) =
    \sqrt{\frac{\tfrac{1}{2} D_{\text{KL}}\!\bigl(P \,\big\|\, \tfrac{P+O}{2}\bigr)
               + \tfrac{1}{2} D_{\text{KL}}\!\bigl(O \,\big\|\, \tfrac{P+O}{2}\bigr)}{\ln 2}},
\end{equation}
and is bounded in $[0, 1]$. 


Since raw distance values are difficult to interpret in isolation, we provide a universally comparable,
chance-corrected alignment score by calibrating against a permutation baseline. 
Given distance $d(P, O)$, we estimate the expected distance under
random label assignment by repeatedly shuffling $O$:
\begin{equation}\label{eq:permutation-baseline}
  \hat{\mathbb{E}}_\pi[d(P, \pi(O))]
    = \frac{1}{T} \sum_{t=1}^{T} d(P, \pi_t(O))
    \qquad (T = 50{,}000)
\end{equation}
The alignment score is then:
\begin{equation}\label{eq:alignment}
  a(P, O) = \max\!\left(0,\;
    1 - \frac{d(P, O)}{\hat{\mathbb{E}}_\pi[d(P, \pi(O))]}
  \right)
\end{equation}

Identical distributions receive $a = 1$; $a = 0$ means the
evaluation is no closer to production than a random reshuffling of
probability mass. Holding $P$ fixed and shuffling only $O$ preserves
$P$'s real-world sparsity as the reference structure, asking ``how
much better than chance is $O$ at reproducing $P$?'' 
An earlier design that drew $O$ from a Dirichlet distribution (always dense) severely
underestimated the baseline distance for sparse production
distributions, yielding misleadingly low alignment scores.


When multiple dimensions are evaluated, each dimension $i$ yields its
own alignment score $a_i$.  To produce a single overall score, we use
a \emph{redundancy-discounted weighted average} that accounts for three
factors: (1)~importance weight $w_i$; (2)~entropy $H_i$ of the
reference distribution; and (3)~pairwise mutual information
$\MI(i,j)$ between dimensions.  The importance weights $w_i$ are set
by domain experts and encode the intuition that not all dimensions
matter equally for downstream quality (for example, drift in user
intent is far more consequential than drift in surface formality).

Dimensions are processed in decreasing order of $w_i \cdot H_i$.
For each dimension $i$, the redundancy discount is:
\begin{align}
  \rho_{ij} &= \MI(i,j)/{H_i}
    & &\text{(overlap with earlier dim $j$)} \\
  r_i &= \sum_{j \prec i} \rho_{ij}
    & &\text{(total redundancy)} \\
  g_i &= \max(0,\; 1 - \lambda \, r_i)
    & &\text{(discount, $\lambda{=}0.8$)}
\end{align}
The effective dimension weight is then set to $\tilde{w}_i = g_i \cdot w_i$, and the
redundancy-aware (RA) alignment is:
\begin{equation}\label{eq:ra-score}
  A_\text{RA} = \frac{\sum_i \tilde{w}_i \cdot a_i}
                      {\sum_i \tilde{w}_i}
\end{equation}

This formulation ensures that highly correlated dimensions are progressively down-weighted, preventing
a cluster of redundant dimensions from dominating the aggregate score.
When all dimensions are independent, it reduces to an ordinary weighted
average. 

Because both $a_i$ and $A_\text{RA}$ are chance-corrected on
$[0, 1]$, raw values can be turned into qualitative labels by partitioning
the range; we use equal thirds (\textcolor{driftbad}{\textbf{bad}}~$<1/3$,
\textcolor{driftaverage}{\textbf{average}}~$\geq 1/3$,
\textcolor{driftgood}{\textbf{good}}~$\geq 2/3$) uniformly
throughout the paper.

%% file: sections/sampling.tex
\section{Synthetic Dataset Generation}\label{sec:sampling}
Data generation is a key component of the \sysname that can be used 
to augment existing eyes-on datasets with identified drift or to generate a new production-aligned one.
Since reusing real production traffic verbatim is not an option due to privacy concerns, and
hand-crafting new sets is slow and inherits the blind spots that
motivated this work, our solution is to generate a synthetic set
that matches production at the level of \emph{distribution
statistics over proxy dimensions}. This method was chosen 
over mirroring individual production queries to avoid overtuning.
The pipeline proceeds in four stages: 
(1)~as a part of production workflow, 
user queries are classified into structured proxy objects and saved 
along with their observed properties in telemetry; 
(2)~proxies are aggregated into per-dimension marginals and pairwise co-occurrence counts; 
(3)~the conditional sampler draws synthetic proxies from these statistics; 
(4)~an LLM generator hydrates each proxy into a natural-language query; 
The remainder of this section details the Chow--Liu tree model
that we used for sampling and two finite-sample corrections
that account for limited data.


A baseline \emph{independence sampler} treats each dimension
independently, drawing from per-dimension marginals.  This discards
all correlations and routinely produces implausible combinations
(e.g., \dimval{Legal Document} output type with a \dimval{Casual} tone) -- artefacts
that exercise the model on inputs no real user would issue, biasing
quality estimates and wasting evaluation budget.  We therefore need
a sampler that respects the joint structure of the dimensions.

We model the joint distribution over $D$ dimensions using the Chow-Liu optimal tree approximation~\cite{chow1968}: the maximum-weight spanning tree over dimensions with mutual information as edge weights. For every dimension pair $(X_i, X_j)$, with vocabularies $\mathcal{V}_i$ and $\mathcal{V}_i$ we maintain a contingency
table of co-occurrence counts from raw proxy records.  For
multi-valued dimensions, every combination of values from the two
dimensions in the same record contributes a count.  Joint-derived
marginals ensure consistency:
\begin{equation}
  p(v_i) = \sum_{v_j} p(v_i, v_j), \quad
  \MI(X_i; X_j) = \sum_{v_i, v_j} p(v_i, v_j)
    \ln \frac{p(v_i, v_j)}{p(v_i)\,p(v_j)}
\end{equation}

The Chow-Liu theorem states that the optimal tree-structured
approximation is the maximum-weight spanning tree with MI as edge
weights. We construct it via Kruskal's algorithm with union-find
(rank compression and path halving), yielding $O(D^2 \log D)$
complexity.

The undirected tree is then rooted at the node with highest degree: the
``hub'' that directly conditions the most children, providing a
short path to a large fraction of the dimensions and limiting the
number of conditional steps over which any single estimation error
can propagate.  The rooted tree defines:
\begin{equation}\label{eq:factorization}
  P(X_1, \ldots, X_D) = P(X_r) \prod_{d \neq r} P(X_d \mid X_{\pi(d)})
\end{equation}
where $r$ is the root and $\pi(d)$ the parent of dimension $d$. 
See an example tree in Figure~\ref{fig:chowliu-tree}.

\input{figures/chowliu_tree}


The textbook Chow--Liu construction assumes that both the pairwise
MI values and the per-parent conditionals can be read off the data
directly.  In practice, neither is true: many dimension pairs are
supported by a modest number of co-occurring records, and many
parent values are themselves rare, so plug-in estimates are noisy in
opposite ways: MI is over-estimated, while conditionals are often
under-determined.  We address each with a tailored correction.

With $N_{ij}$ as the number of co-occurring records for the pair $(X_i, X_j)$, MI computed by plug-in from finite samples exhibits a positive
bias~\cite{brillinger2004} scaling as
$(|\mathcal{V}_i|{-}1)(|\mathcal{V}_j|{-}1) / (2N_{ij})$.  Left
uncorrected, this bias inflates the apparent dependence between
high-cardinality dimensions and can promote spurious edges into the
maximum spanning tree. Rather than subtracting an analytical
correction that can overshoot and yield negative MI estimates, we
apply a multiplicative \emph{sigmoid dampening}:
\begin{equation}\label{eq:dampen}
  \widehat{\MI}(X_i; X_j) = \MI(X_i; X_j) \cdot
    \sigma\!\left(
      \ln \frac{N_{ij}}{c \cdot |\mathcal{V}_i| \cdot |\mathcal{V}_j|}
    \right),
\end{equation}
where $\sigma$ is the logistic sigmoid and $c=5$ follows Cochran's
rule~\cite{cochran1954} for minimum expected cell counts.  The
factor smoothly suppresses MI when sample size is small relative to
the table area $|\mathcal{V}_i| \cdot |\mathcal{V}_j|$ and approaches $1$
when evidence is plentiful.  By construction it preserves zeros
($\hat{I} = 0$ when $I = 0$), is strictly less than $I$ for any
finite sample (asymptotically approaching $I$ as $N_{ij} \to
\infty$), and adapts automatically across pairs of differing
cardinality.

Once the tree is fixed, each non-root dimension $X_c$ must be
sampled from $P(X_c \mid X_p = v_p)$ for every value $v_p$ of its
parent. Parent values vary widely in support: rare ones yield
near-degenerate empirical conditionals that lock in spurious
dependencies the data do not actually justify. We therefore \emph{shrink each
conditional toward the marginal prior} $P_0(X_c)$ by a coefficient
that reflects how strongly the data itself argues for a difference from the prior:

\textbf{Step 1:} We start with the empirical conditional:
\begin{equation}
\hat{P}(X_c {=} v_c \mid X_p {=} v_p) = c(v_p, v_c) / n_{v_p}.
\end{equation}

\textbf{Step 2:} We test whether the empirical conditional differs from the marginal
prior $P_0(X_c)$ using Pearson's $\chi^2$:
\begin{equation}
  \chi^2 = \sum_{v_c}
    \frac{(O_{v_c} - E_{v_c})^2}{E_{v_c}}, \quad
  E_{v_c} = n_{v_p} \cdot P_0(v_c).
\end{equation}
We apply a relaxed variant of Cochran's rule~\cite{cochran1954},
requiring at least one expected count per cell on average
($n_{v_p} \geq |\mathcal{V}_c|$); when this fails the $\chi^2$
approximation is unusable and we set $\gamma = 1$ (pure prior).

\textbf{Step 3:} The shrinkage coefficient is then set to:
\begin{equation}\label{eq:shrinkage}
  \gamma_{v_p} = \sqrt{p\text{-value}}.
\end{equation}
The square root softens the transition: significant conditionals
($p~\ll~1$) keep $\gamma \approx 0$ and trust the data, borderline
cases ($p \approx 0.1$) receive moderate shrinkage, and
non-significant slices ($p \to 1$) collapse back to the prior.

\textbf{Step 4:} The final interpolated conditional:
\begin{equation}\label{eq:interpolate}
  P(X_c {=} v_c \mid X_p {=} v_p) =
    (1 - \gamma_{v_p})\,\hat{P}(v_c \mid v_p)
    + \gamma_{v_p}\,P_0(v_c).
\end{equation}

Together, the two described corrections ensure that the sampler only commits
to a dependency (either as an edge in the tree or as a sharp
conditional within an edge) when the data demonstrably supports it,
and otherwise falls back to the prior estimate.


Given the fitted model above, sampling follows the tree factorization in
topological order: sample the root from its prior, then each child
from its conditional given its parent's sampled value.  Each sample
produces a complete proxy object in $O(D)$ time.

The sampled proxy is then converted into a natural-language query by
an LLM generator. The generator de-facto reverses the work of proxification: it receives the proxy as structured
input with the same schema as the LLM classifier (except for observable properties) and produces a natural language query. 

%% file: figures/chowliu_tree.tex
\begin{figure}[t]
\centering
\includegraphics[width=0.9\columnwidth]{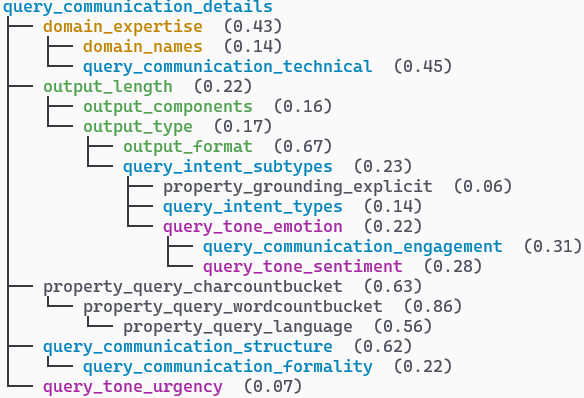}
\caption{Chow--Liu maximum spanning tree for one application
  scenario, built from the initial dimension taxonomy (prior to
  the refinement described in \S\ref{sec:eval-refinement}).
  Edge labels are sigmoid-dampened mutual information
  (\S\ref{sec:sampling}).
  Color indicates semantic category:
  \textcolor{catintent}{\textbf{intent \& style}},
  \textcolor{cattone}{\textbf{tone \& affect}},
  \textcolor{catoutput}{\textbf{output}},
  \textcolor{catcontent}{\textbf{content}},
  \textcolor{catobs}{\textbf{observable}}.}
\label{fig:chowliu-tree}
\end{figure}

%% file: sections/deployment.tex
\section{System Deployment and Data}\label{sec:deployment}

This section describes the production system in which \sysname
operates and the datasets used in the evaluation that follows.


\sysname is deployed within a large-scale agentic chat system that
serves hundreds of millions of users across a cloud-based
productivity suite, orchestrating multiple specialized
scenarios that include generating, editing documents 
(presentations, reports, spreadsheets, etc.) 
on the user's behalf. These two skills are the focus of the present work. 
Within the production compliance boundary, a dedicated workflow
fires after each session: the LLM classifier
(\S\ref{sec:proxy}) emits a proxy object that is persisted
in telemetry -- the only representation of user interactions that
leaves the live environment.  Scheduled pipelines aggregate proxies
\emph{daily} into per-dimension marginals and pairwise co-occurrence
tables, one set per scenario; these aggregates feed both the drift
detector (\S\ref{sec:drift}) and the synthetic data generator
(\S\ref{sec:sampling}).


The evaluation in \S\ref{sec:eval} draws on multiple datasets:
(0)~\emph{legacy hand-curated sets}, one per artifact-type
scenario, assembled over the product's lifetime for functional
coverage rather than distributional fidelity. In \S\ref{sec:eval-alignment} they are used as the
baseline against which synthetic datasets are compared;
(1)~a \emph{designed synthetic set} of 450~queries that is 
representative of production marginals and covers all dimension
values, including rare ones, from our taxonomy. It is used in the roundtrip experiments
(\S\ref{sec:eval-roundtrip}--\S\ref{sec:eval-refinement});
(2)~two \emph{eyes-on user-feedback sets} with production interactions and binary thumbs-up/thumbs-down labels. (2a) for document generation: 5.3K interactions (70.5\% positive; used in \S\ref{sec:eval-dimpower}) and (2b) for comprehension: 2.3K interactions (26.8\% positive; used in \S\ref{sec:eval-case}); and
(3)~a balanced 200-query \emph{discrimination set} (100 from
dataset~(1) and 100 from dataset~(2a)) used in
\S\ref{sec:eval-naturalness}.



%% file: sections/evaluation.tex
\section{Evaluation}\label{sec:eval}

We evaluate \sysname along four complementary axes:
(1)~end-to-end drift alignment between legacy/synthetic and production
proxy distributions; (2)~generation+classification roundtrip consistency of the proxy
representation and its use for taxonomy refinement;
(3)~indistinguishability of synthetic queries from human queries;
and (4)~discriminative power of proxy dimensions for predicting
user satisfaction.  We deliberately discuss alignment first because
it motivated the schema-refinement loop in (2): the very gap
between the legacy offline test set and production traffic is what
prompted us to build the (conditional) sampler whose proxies are then
put through the roundtrip experiments.

\subsection{End-to-End Drift Alignment}\label{sec:eval-alignment}

This experiment measures whether the full \sysname pipeline can
produce a synthetic dataset whose proxy distribution matches
production.  We use the redundancy-aware (RA) alignment score of
\S\ref{sec:drift} and report results for the Excel
artifact-creation scenario as the results for other document types (e.g., PowerPoint)
are almost identical. This evaluation predates the
taxonomy-refinement loop of \S\ref{sec:eval-refinement}:
even against the unrefined schema, it shows that distribution-aware
sampling is worth pursuing in the first place.

\paragraph{Setup.}
Reference distributions were computed from a week of production traffic (2026-04-02 to 2026-04-08).  Two samplers each produced a synthetic dataset of ${\sim}1{,}000$ proxies: the \emph{basic} (dimensionally independent) sampler, drawing each dimension from its production marginal, and the \emph{conditional}
(Chow-Liu) sampler of \S\ref{sec:sampling}.  As a third point of
reference we re-scored the pre-existing hand-curated Excel test set
(\S\ref{sec:deployment}, dataset~(0)) against the same production baseline.

\begin{table}[t]
\centering
\caption{RA alignment (\S\ref{sec:drift}) for the Excel
  document generation scenario.}
\label{tab:ra-scores}
\small
\begin{tabular}{@{}lcc@{}}
\toprule
\textbf{Dataset} & \textbf{RA score} & \textbf{Band} \\
\midrule
Legacy hand-curated set         & 0.284    & \textcolor{driftbad}{\textbf{bad}}     \\
Basic (independent) sampler     & 0.897  & \textcolor{driftgood}{\textbf{good}}    \\
Conditional (Chow-Liu) sampler & 0.917  & \textcolor{driftgood}{\textbf{good}}    \\
\bottomrule
\end{tabular}
\end{table}


\paragraph{Results.}
As Table~\ref{tab:ra-scores} shows, both synthetic samplers land
squarely in the \emph{good} band (cf. \S\ref{sec:drift}), with the conditional sampler
outperforming the basic one (RA $0.917$ vs.\ $0.897$).  The basic sampler scores high because the RA metric
evaluates per-dimension marginals and the basic sampler is
calibrated directly to them; the conditional sampler edges past it
by additionally preserving cross-dimension structure (e.g.,
intent-output co-occurrences) without sacrificing marginal
fidelity.  By contrast, the legacy hand-curated test set fails on
most dimensions, with individual per-dimension alignment scores typically
below $0.40$ and the overall alignment score in the \emph{bad} range.
This gap also impacts evaluation outcomes: in the Excel agent scenario, 
scores on the distribution-aligned synthetic dataset are consistently
lower than on the legacy offline set (e.g., Accuracy 4.29$\to$3.71, Reliability 3.85$\to$3.44; avg.\ $\sim$0.3 decrease).\footnote{Each (query, generated document) pair is scored by an
LLM-as-a-judge~\cite{zheng2023judging,kim2024prometheus2} on a 1--5
scale across five rubric dimensions, including Accuracy (correctness
of query interpretation and implementation), Completeness (coverage of requirements), 
Usefulness (practical value for user goals), Reliability (technical consistency and robustness), 
and Aesthetics (formatting and presentation quality).} 
This suggests systematic score inflation in conventional offline evaluation, 
directly supporting the decision distortion described in the introduction.
This gap is the practical justification for the rest of the
pipeline: distribution-aware synthetic data closes a quality gap
that manual curation cannot.

\subsection{Roundtrip Consistency}\label{sec:eval-roundtrip}

The roundtrip experiment investigates internal consistency of the proxy representation: 
does a proxy, when used to generate a
synthetic query and then reclassified, reconstruct itself?
We designed the experiment with two purposes in mind: as a
diagnostic that quantifies how faithfully the
generation-classification cycle preserves proxy information, and as
the optimization signal that drives the taxonomy refinement loop of
\S\ref{sec:eval-refinement}: each iteration of the schema and
prompts is judged by the reduction in roundtrip distance and in the
per-dimension confusion patterns it produces.

\paragraph{Protocol.}
The $M$ input proxies are obtained by running the LLM classifier on
the designed synthetic dataset (\S\ref{sec:deployment}, dataset~(1))
and joining each predicted proxy with the observable properties
measured directly from the corresponding query (language, word-length buckets, explicit grounding markers).  For each
of these input proxies: (1)~the proxy is given to the LLM
generator, which produces $N = 3$ synthetic queries; (2)~each query
is independently classified back into a proxy (with its observable
properties re-measured); (3)~for each (original, reconstructed) pair,
a per-dimension distance is computed and aggregated into a scalar
$d \in [0,1]$.  Per-dimension distances are confidence-weighted and
account for value identity, ordinal proximity (for sorted dimensions),
and -- for multi-valued dimensions -- set overlap.

\paragraph{Initial results.}
The experiment was first conducted against an initial,
intuition-driven taxonomy on the $M=450$ input proxies derived from
dataset~(1).  Of these, 31 (6.9\%) failed to complete the
generation-classification cycle typically due to the LLM
classifier failure to produce a schema-compliant output for the
generated query within the self-correction budget. For the 
remaining 419 roundtrips (3 generations each), the mean
reconstruction distance was $\bar{\mu} = 0.291$, with the
within-proxy standard deviation ($\bar\sigma = 0.055$)
substantially smaller than the across-proxy standard deviation
($s_\mu = 0.1$).  This pattern revealed that the residual
mismatch was \emph{systematic} rather than random: specific
dimension values were consistently confused, pointing to actionable
taxonomy deficiencies rather than general noise.

\subsection{Taxonomy Refinement via Confusion Analysis}\label{sec:eval-refinement}

By computing per-dimension \emph{confusion matrices} from the
roundtrip outputs, we identified which dimensions and values were
conflated during the cycle.  Each matrix is built by
proportional-redistribution mass accounting that preserves both
original and reconstructed marginals, and is summarized by a
size-adjusted \emph{diagonal-concentration} score: the share of
mass on (or, for ordinal dimensions, near) the main diagonal,
rescaled to compensate for matrix size.  This analysis directly
guided iterative improvements to the proxy schema. We:

\begin{itemize}
  \item \textbf{retired dimensions} that carried little signal and
    were consistently confused (e.g., domain expertise,
    intent categories);
  \item \textbf{revised value sets}, i.e., merged values too similar for
    the LLM to discriminate and split overly broad categories;
  \item \textbf{introduced new dimensions} for constructs that were
    conflated within existing ones (e.g., separating
    content specificity from communication constraints);
  \item \textbf{rewrote definitions and examples} in both classifier
    and generator prompts, adding disambiguation guidance for
    borderline cases identified from the confusion matrices.
\end{itemize}

\noindent
The resulting taxonomy (the post-optimization schema) organizes
dimensions into the categories shown in
Table~\ref{tab:schema-overview}.

\input{figures/schema_overview}
\begin{table}[t]
\centering
\caption{Roundtrip consistency before and after taxonomy refinement and for the two
  multi-turn variants ($M=450$ input proxies, $N{=}3$ generations each)
  Lower $\bar{\mu}$ is better.}
\label{tab:roundtrip}
\small
\setlength{\tabcolsep}{4pt}
\begin{tabular}{@{}lrrrr@{}}
\toprule
& \multicolumn{2}{c}{\textbf{Single-turn}} & \multicolumn{2}{c}{\textbf{Multi-turn}}\\
\cmidrule(lr){2-3}\cmidrule(lr){4-5}
\textbf{Statistic} & \textbf{Before} & \textbf{After} & \textbf{Inter.} & \textbf{Query} \\
\midrule
Failed roundtrips                    & 31    & 2     & 4     & 3     \\
Mean proxy distance ($\bar{\mu}$)    & 0.282 & 0.158 & 0.186 & 0.153 \\
Std.\ dev.\ of means ($s_\mu$)       & 0.100 & 0.076 & 0.084 & 0.085 \\
Mean within-proxy std ($\bar\sigma$) & 0.055 & 0.042 & 0.046 & 0.041 \\
\bottomrule
\end{tabular}
\end{table}

Table~\ref{tab:roundtrip} shows the aggregate roundtrip statistics
before and after taxonomy refinement, on the same set of
450~input proxies.
Refinement cut the mean reconstruction distance from 0.282 to
0.158 (a 44\% relative reduction) and dropped the roundtrip failure
rate from 6.9\% to 0.4\%: evidence
that the revised schema is far easier for the LLM to apply
consistently.  Formerly problematic dimensions also improved
sharply (\dimname{query\_engagement}: 0.51\,$\to$\,0.81 macro
diagonal concentration; \dimname{query\_structure}:
0.65\,$\to$\,0.81), while already-strong dimensions held
(\dimname{output\_type}: 0.89\,$\to$\,0.95).

\input{figures/confusion_combined}

Figure~\ref{fig:confusion-matrices} shows representative confusion
matrices for two dimension archetypes:
\dimname{tone\_emotion} (Fig.~\ref{fig:conf-tone-emotion}), a
\emph{multi-valued} dimension capturing emotional tone, in which
high-frequency emotions reconstruct reliably while rarer ones
(\dimval{Curiosity}, \dimval{Uncertainty}) collapse into \dimval{Neutral}; and
\dimname{query\_constraints} (Fig.~\ref{fig:conf-comm-constraints}),
an \emph{ordinal} dimension whose neighbor-smearing pattern
(\dimval{Minimal} often reconstructed as \dimval{Moderate}) reflects genuine ambiguity
in the underlying construct rather than model failure.  For
comparison, \dimname{output\_type} (\emph{nominal}, 11 values)
achieves macro diagonal concentration 0.95, a near-perfect document-type
recovery.

Encouragingly, Table~\ref{tab:roundtrip} also shows that both multi-turn schema variants
(\S\ref{sec:proxy}) behave consistently with the single-turn
baseline with the
roundtrip yields mean reconstruction distances of
$\bar\mu = 0.186$ at the interaction level and $\bar\mu = 0.153$
at the query level -- both within $0.03$ of the single-turn
$\bar\mu = 0.158$, and per-dimension diagonal-concentration factors
also falling in the same range as their single-turn counterparts.  This
suggests that the proxy representation extends naturally to
multi-turn settings without loss of fidelity.

\subsection{Naturalness Test}\label{sec:eval-naturalness}

To assess naturalness of the synthetic data, we submitted 100 human queries (from production feedback logs) and
100 synthetic queries to an independent LLM discriminator under two
conditions: a \emph{guided} prompt (enumerating known synthetic
tells such as overly polished grammar and balanced structure) and
an \emph{unguided} prompt (verdict only).

\begin{table}[t]
\centering
\caption{Naturalness test: LLM accuracy at distinguishing human
  vs.\ synthetic queries ($n = 200$, chance = 50\%).}
\label{tab:naturalness}
\small
\begin{tabular}{@{}lcccc@{}}
\toprule
\textbf{Condition} & \textbf{Accuracy} & \textbf{$p$-value}
  & \textbf{Human rec.} & \textbf{Synth rec.} \\
\midrule
Guided   & 0.650 & $1.3 \times 10^{-5}$ & 0.91 & 0.39 \\
Unguided & 0.540 & 0.144 & 0.92 & 0.16 \\
\bottomrule
\end{tabular}
\end{table}

As shown in Table~\ref{tab:naturalness}, without explicit guidance
the discriminator cannot reliably distinguish synthetic from human
queries ($p = 0.14$).  With guidance, accuracy rises to 65\%, but
the gain comes entirely from identifying human queries via surface
markers (typos, code-switching, implicit context); synthetic recall
remains low (39\%), meaning most synthetic queries are still
classified as human.

\subsection{Predictive Signal Analysis}\label{sec:eval-dimpower}

Beyond measuring drift, we ask: \emph{which proxy dimensions carry
actionable information about system quality?}  We evaluate this using
the feedback signal dataset~(2a) with binary user
satisfaction (thumbs-up/down) as the target signal $Y$.

For each dimension $X$ we compute the held-out conditional
log-likelihood (CLL) gain of predicting $Y$ from $X$ versus the
marginal under stratified 5-fold cross-validation, with conditionals
estimated under Laplace smoothing~\cite{lidstone1920}.  Ordinal dimensions optionally
use pool-adjacent-violators (PAV) refinement~\cite{zadrozny2002transforming}, with the nested
PAV-vs-raw tiebreaker on the training fold settled by
chance-corrected AMI \cite{vinh2010ami} with Kish's effective-sample-size
correction \cite{kish1965survey}.  We call a dimension's signal
\emph{stable} when its per-fold mean $\bar\mu$ is positive and at
least twice the per-fold standard deviation (SNR $\bar\mu/\sigma \geq 2$).

Applied to 5,293 interactions, seven dimensions
clear the SNR~$\geq 2$ bar: sentiment, intent type, emotion,
domain, output creativity, formality, and query word count, with
CLL gains in $[0.0019, 0.0152]$~nats per fold.  A second cluster
(output type, output format, query language) has positive means
dominated by a handful of rare classes, so per-fold variance is
high; the remaining dimensions carry essentially no signal.

A complementary per-value diagnostic, the lift\\
$\hat P(Y{=}1\mid X{=}v) - \hat P(Y{=}1)$ with significance assessed
by a one-vs-rest binomial $z$-contrast, reveals that dimension-level
scores are diluted by dominant uninformative modes. Individual
values carry stronger signal in both directions, e.g., queries
laced with \dimval{Frustration} elicit thumbs-down 93\% of the time
($z = -7.5$), while \dimval{Summarization} requests lift thumbs-up to 51\%
($z = +5.2$, vs.\ a 27\% baseline).  These per-value scores directly
inform which dimensions to prioritize in drift monitoring and which
production segments to investigate or amplify when quality
shifts.

\subsection{Telemetry-Driven Copilot Improvement for Question Answering (QA) over Excel Files}
\label{sec:eval-case}

The feedback-linkage analysis in \S\ref{sec:eval-dimpower} shows that
proxy dimensions can expose user-facing failure modes without access to
raw user content. We next demonstrate its portability by applying the diagnostic to QA over Excel files in Microsoft Copilot.

Given an Excel workbook and a query, Copilot must answer
by understanding various workbook semantics. Many queries
depend on formulas, filters, sorting state, charts, formatting, or data
validation. We therefore define Excel-specific binary proxy dimensions:
\textsf{Formula=yes}, \textsf{Filter/Sort=yes}, \textsf{Chart=yes},
and \textsf{Cell-Styling=yes}, indicating whether each attribute is
needed for question answering. These proxy values turn
heterogeneous spreadsheet requirements into measurable, actionable
signals.

We instantiate the diagnostic on $2{,}346$ Copilot ask-queries over Excel
files with thumb-up/down feedback (dataset (2b)), where thumbs-down is treated as
dissatisfaction (DSAT) and thumbs-up as satisfaction (SAT). Overall,
$29.5\%$ of queries receive negative feedback (DSAT). The data is
stratified into matched train/test splits of $1{,}174$/$1{,}172$ queries. On the training split, we
apply the per-value binomial $z$-contrast from \S\ref{sec:eval-dimpower},
comparing each \textsf{yes} value against the global DSAT baseline. The
held-out test split is reserved for measuring how many DSAT cases become
satisfactory after refinement. As shown in Table~\ref{tab:excel-case},
all four proxy values have elevated DSAT rates, suggesting that specific workbook semantics challenges Excel QA.

\begin{table}[t]
\centering
\caption{Case study on QA over Excel files. The left block reports
training-split DSAT diagnostics for Excel-specific proxy values. The
right block reports post-refinement gains on sampled held-out telemetry DSAT
cases and on sampled QA pairs from public spreadsheet benchmarks.}
\label{tab:excel-case}
\small
\setlength{\tabcolsep}{3.5pt}
\begin{tabular}{@{}lrrr!{\vrule width 0.6pt}rr@{}}
\toprule
& \multicolumn{3}{c!{\vrule width 0.6pt}}{\textbf{DSAT Diagnostic}}
& \multicolumn{2}{c}{\textbf{Post-refinement Gain}} \\
\cmidrule(lr){2-4} \cmidrule(lr){5-6}
\textbf{Proxy value} & \textbf{$n$} & \textbf{DSAT} & \textbf{$z$}
& \textbf{DSAT$\to$SAT} & \textbf{Public Acc.} \\
\midrule
Filter/Sort = yes   &  43 & 41.9\% & $+1.77$ & 40.0\% & $+50.0$ pts \\
Formula = yes       & 176 & 34.7\% & $+1.49$ & 60.0\% & $+55.6$ pts \\
Chart = yes         & 110 & 34.5\% & $+1.15$ & 20.0\% & $+16.7$ pts \\
Cell-Styling = yes  &  26 & 34.6\% & $+0.57$ & 40.0\% & $+33.4$ pts \\
\bottomrule
\end{tabular}
\end{table}

We used this signal to refine the Excel encoding consumed by Copilot. Inspection showed that production Excel encoding dropped
specific semantics to save context length. We augmented the encoding to preserve formulas, sort/filter state, data validation, and
merged-cell structure, making these properties explicit to the model.

For held-out telemetry validation, we then
sample five DSAT cases per proxy category and assess whether the refined 
encoding flips DSAT outcomes to SAT. The refinement recovers $60.0\%$,
$40.0\%$, $20.0\%$, and $40.0\%$ of Formula, Filter/Sort, Chart, and
Cell-Styling cases, respectively, illustrating how ProxyDrift turns
privacy-preserving proxy diagnostics into engineering improvements for
Microsoft Copilot. It also yields large gains on public spreadsheet benchmarks sampled
to match the telemetry feature distribution, including
RealHiTBench~\cite{wu2025realhitbench}, SheetBench~\cite{wang2026sheetbrain},
and MiMoTable~\cite{limimotable}. Accuracy improves by $+55.6$, $+50.0$,
$+16.7$, and $+33.4$ points on Formula, Filter/Sort, Chart, and
Cell-Styling queries, respectively. 

%% file: figures/schema_overview.tex
\begin{table}[t]
\centering
\small
\setlength{\tabcolsep}{3pt}
\renewcommand{\arraystretch}{1.05}
\begin{tabular}{@{}llccc@{}}
\toprule
\textbf{Category} & \textbf{Dimension} & $|\mathcal{V}|$ & \textbf{Type} & \textbf{Ord.} \\
\midrule
\multirow{6}{*}{\textcolor{catintent}{\romark{Intent \& Style}}}
& \textcolor{catintent}{\dimname{intent\_types}}           & 25 & multi & --- \\
& \textcolor{catintent}{\dimname{query\_formality}}        &  4 & single & \checkmark \\
& \textcolor{catintent}{\dimname{query\_structure}}        &  4 & single & --- \\
& \textcolor{catintent}{\dimname{query\_technical}}        &  4 & single & \checkmark \\
& \textcolor{catintent}{\dimname{query\_engagement}}       &  5 & multi & --- \\
& \textcolor{catintent}{\dimname{query\_constraints}}      &  5 & single & \checkmark \\
\midrule
\multirow{3}{*}{\textcolor{cattone}{\romark{Tone \& Affect}}}
& \textcolor{cattone}{\dimname{tone\_sentiment}}         &  6 & single & \checkmark \\
& \textcolor{cattone}{\dimname{tone\_emotion}}           &  9 & multi & --- \\
& \textcolor{cattone}{\dimname{tone\_urgency}}           &  5 & single & \checkmark \\
\midrule
\multirow{5}{*}{\textcolor{catoutput}{\romark{Output}}}
& \textcolor{catoutput}{\dimname{output\_type}}            & 11 & single & --- \\
& \textcolor{catoutput}{\dimname{output\_format}}          &  6 & single & --- \\
& \textcolor{catoutput}{\dimname{output\_length}}          &  5 & single & \checkmark \\
& \textcolor{catoutput}{\dimname{output\_components}}      &  6 & multi & --- \\
& \textcolor{catoutput}{\dimname{output\_creativity}}      &  4 & single & \checkmark \\
\midrule
\multirow{3}{*}{\textcolor{catcontent}{\romark{Content}}}
& \textcolor{catcontent}{\dimname{domain\_names}}           & 21 & multi & --- \\
& \textcolor{catcontent}{\dimname{content\_specificity}}    &  4 & single & \checkmark \\
& \textcolor{catcontent}{\dimname{content\_embedded}}       &  5 & multi & --- \\
\midrule
\multirow{4}{*}{\textcolor{catobs}{\romark{Observable}}}
& \textcolor{catobs}{\dimname{char\_count\_bucket}}     &  6 & single & \checkmark \\
& \textcolor{catobs}{\dimname{word\_count\_bucket}}     &  6 & single & \checkmark \\
& \textcolor{catobs}{\dimname{query\_language}}         & 11 & single & --- \\
& \textcolor{catobs}{\dimname{grounding\_explicit}}     & 20 & multi & --- \\
\bottomrule
\end{tabular}
\caption{Post-optimization proxy schema (single-turn).
  $D = 21$ dimensions: 17 LLM-classified and 4 observable.
  $|\mathcal{V}|$ is vocabulary size.
  ``Ord.'' stands for ordinal (has a meaningful value ordering).}
\label{tab:schema-overview}
\end{table}

%% file: figures/confusion_combined.tex
\begin{figure}[t]
\centering
\setlength{\tabcolsep}{1.2pt}
\renewcommand{\arraystretch}{1.1}
\scriptsize

\begin{subfigure}[t]{\columnwidth}
\centering
\input{figures/confusion_tone_emotion}
\caption{\dimname{tone\_emotion} (multi-valued, 8~values).
  Macro conc.: 0.79.}
\label{fig:conf-tone-emotion}
\end{subfigure}

\vspace{4pt}

\begin{subfigure}[t]{\columnwidth}
\centering
\input{figures/confusion_comm_constraints}
\caption{\dimname{query\_constraints} (ordinal, 5~values).
  Macro conc.: 0.85.}
\label{fig:conf-comm-constraints}
\end{subfigure}

\caption{Row-normalized confusion matrices for two representative
  dimensions. 
  (a)~\textbf{Multi-valued}: common emotions recovered well; rare
  ones collapse into \dimval{Neutral}.
  (b)~\textbf{Ordinal}: neighbor-smearing (\dimval{Minimal}
  $\leftrightarrow$ \dimval{Moderate}) reflects genuine construct ambiguity.
  Cf.\ \dimname{output\_type} (nominal, macro conc.\ 0.95) for the
  high-diagonal nominal case.}
\label{fig:confusion-matrices}
\end{figure}

%% file: figures/confusion_tone_emotion.tex
\begin{tabular}{@{}r r|cccccccc@{}}
  & \scriptsize$n$ & \romark{Enthusiasm} & \romark{Curiosity} & \romark{Frustration} & \romark{Impatience} & \romark{Confusion} & \romark{Uncertainty} & \romark{Neutral} & \romark{Other} \\
  \hline
  \footnotesize Enthusiasm & {\scriptsize 75} & \cellcolor{confblue!70}\textcolor{white}{\footnotesize 71\%} &  &  &  &  &  & \cellcolor{confblue!29}\textcolor{black}{\footnotesize 29\%} &  \\
  \footnotesize Curiosity & {\scriptsize 17} &  & \cellcolor{confblue!35}\textcolor{black}{\footnotesize 35\%} &  &  &  &  & \cellcolor{confblue!64}\textcolor{white}{\footnotesize 65\%} &  \\
  \footnotesize Frustration & {\scriptsize 77} &  &  & \cellcolor{confblue!75}\textcolor{white}{\footnotesize 75\%} & \cellcolor{confblue!5} & \cellcolor{confblue!7}\textcolor{black}{\footnotesize 8\%} & \cellcolor{confblue!5} & \cellcolor{confblue!12}\textcolor{black}{\footnotesize 13\%} &  \\
  \footnotesize Impatience & {\scriptsize 9} &  &  &  & \cellcolor{confblue!11}\textcolor{black}{\footnotesize 11\%} &  &  & \cellcolor{confblue!88}\textcolor{white}{\footnotesize 89\%} &  \\
  \footnotesize Confusion & {\scriptsize 27} &  & \cellcolor{confblue!5} &  &  & \cellcolor{confblue!77}\textcolor{white}{\footnotesize 78\%} & \cellcolor{confblue!5} & \cellcolor{confblue!14}\textcolor{black}{\footnotesize 15\%} &  \\
  \footnotesize Uncertainty & {\scriptsize 33} &  & \cellcolor{confblue!21}\textcolor{black}{\footnotesize 21\%} & \cellcolor{confblue!5} &  & \cellcolor{confblue!9}\textcolor{black}{\footnotesize 9\%} & \cellcolor{confblue!30}\textcolor{black}{\footnotesize 30\%} & \cellcolor{confblue!36}\textcolor{black}{\footnotesize 36\%} &  \\
  \footnotesize Neutral & {\scriptsize 1087} & \cellcolor{confblue!5} & \cellcolor{confblue!5} &  &  & \cellcolor{confblue!5} & \cellcolor{confblue!5} & \cellcolor{confblue!95}\textcolor{white}{\footnotesize 95\%} &  \\
  \footnotesize Other & {\scriptsize 6} &  &  & \cellcolor{confblue!83}\textcolor{white}{\footnotesize 83\%} &  &  & \cellcolor{confblue!16}\textcolor{black}{\footnotesize 17\%} &  &  \\
\end{tabular}

%% file: figures/confusion_comm_constraints.tex
\begin{tabular}{@{}r r|ccccc@{}}
  & \scriptsize$n$ & \romark{Unknown} & \romark{Minimal} & \romark{Moderate} & \romark{Detailed} & \romark{Comprehensive} \\
  \hline
  \footnotesize Unknown & {\scriptsize 9} &  &  & \cellcolor{confblue!88}\textcolor{white}{\footnotesize 89\%} & \cellcolor{confblue!11}\textcolor{black}{\footnotesize 11\%} &  \\
  \footnotesize Minimal & {\scriptsize 589} & \cellcolor{confblue!5} & \cellcolor{confblue!35}\textcolor{black}{\footnotesize 35\%} & \cellcolor{confblue!55}\textcolor{white}{\footnotesize 56\%} & \cellcolor{confblue!8}\textcolor{black}{\footnotesize 8\%} &  \\
  \footnotesize Moderate & {\scriptsize 575} &  & \cellcolor{confblue!5} & \cellcolor{confblue!58}\textcolor{white}{\footnotesize 58\%} & \cellcolor{confblue!39}\textcolor{black}{\footnotesize 40\%} &  \\
  \footnotesize Detailed & {\scriptsize 131} &  & \cellcolor{confblue!5} & \cellcolor{confblue!5} & \cellcolor{confblue!89}\textcolor{white}{\footnotesize 89\%} & \cellcolor{confblue!5}\textcolor{black}{\footnotesize 5\%} \\
  \footnotesize Comprehensive & {\scriptsize 27} &  &  &  & \cellcolor{confblue!7}\textcolor{black}{\footnotesize 7\%} & \cellcolor{confblue!92}\textcolor{white}{\footnotesize 93\%} \\
\end{tabular}

%% file: sections/conclusion.tex
\section{Conclusion}\label{sec:conclusion}

We presented \sysname, a privacy-preserving framework for detecting,
measuring, and mitigating data drift in large-scale LLM applications.
Operating entirely on structured proxy representations, the system
enables continuous distributional monitoring under strict compliance
constraints via a permutation-calibrated alignment score and a
Chow--Liu conditional sampler with adaptive chi-squared shrinkage
that respects inter-dimensional dependencies.  A production
deployment serving hundreds of millions of users demonstrates high
roundtrip consistency, synthetic queries an unguided LLM
discriminator cannot reliably distinguish from human queries, strong
end-to-end alignment (RA $\approx 0.9$), and proxy dimensions that
carry measurable signal about user satisfaction.  Future work
includes extending the framework to grounding documents and
generated artifacts, correlating proxy dimensions with runtime
quality metrics, 
and using synthetic data for agent optimization beyond evaluation. 

%% file: sections/appendix.tex
\appendix
\section{Proxy Schema: Full Dimension Reference}
\label{sec:appendix-schema}

Table~\ref{tab:schema-reference} expands the compact summary of
Table~\ref{tab:schema-overview} into a full reference for the
post-optimization proxy schema (single-turn): $D = 21$
dimensions across five categories, with 17 LLM-classified and 4
directly observable dimensions.  In the \textbf{Properties} column, ``single''~/~``multi''
indicates whether the dimension carries one value or a set,
``ord.''~/~``nom.'' distinguishes ordinal from nominal value sets,
and $|\mathcal V|$ counts all admissible values (including
\dimval{Unknown} for single-valued dimensions; multi-valued
dimensions encode \dimval{Unknown} as the empty set).  Each
classified value carries a $1$--$5$ confidence; observable
dimensions always carry confidence $s = 5$ when known.  In the
\textbf{Values} column, ordinal value lists are written in rank
order (low~$\to$~high).  The multi-turn schema variants used in
\S\ref{sec:eval-roundtrip}--\S\ref{sec:eval-refinement} share the
same categories and definitions; they add a small number of
turn-level dimensions whose value sets are direct generalizations
of those listed here.

\renewcommand{\arraystretch}{1.15}
\setlength{\LTpre}{6pt}
\setlength{\LTpost}{6pt}
{\small
\setlength{\tabcolsep}{4pt}
\begin{longtable}{@{}
  >{\raggedright\arraybackslash}p{1.15cm}
  >{\raggedright\arraybackslash}p{2.85cm}
  >{\raggedright\arraybackslash}p{3.25cm}
  >{\raggedright\arraybackslash}p{1.5cm}
  >{\raggedright\arraybackslash}p{4.1cm}@{}}
\caption{Post-optimization proxy schema (single-turn):
full description and value sets for all $D = 21$ dimensions,
organized by the five categories of
Table~\ref{tab:schema-overview}.}
\label{tab:schema-reference}\\
\toprule
\textbf{Group} & \textbf{Dimension} & \textbf{Description}
  & \textbf{Properties} & \textbf{Values} \\
\midrule
\endfirsthead
\multicolumn{5}{@{}l}{\textit{Table~\ref{tab:schema-reference}
  (continued).}}\\
\toprule
\textbf{Group} & \textbf{Dimension} & \textbf{Description}
  & \textbf{Properties} & \textbf{Values} \\
\midrule
\endhead
\midrule
\multicolumn{5}{r@{}}{\textit{(continued on next page)}}\\
\endfoot
\bottomrule
\endlastfoot
\multirow{6}{=}{\textit{Intent \& Style}}
  & \dimname{intent\_types}
  & What action the user is requesting; multiple intents may
    co-occur in a single query.
  & multi, nom., $|\mathcal V|=25$
  & Factual Inquiry; Conceptual Explanation; Procedural Guidance;
    Definitional Query; Comparative Analysis; Recommendation
    Seeking; Troubleshooting; Verification; Task Execution; Content
    Generation; Document Creation; Document Edit; Template
    Generation; Structured Output; Code Generation; Data
    Transformation; Summarization; QnA; Research; Greeting;
    Clarification Request; Follow Up Question; Scope Modification;
    Capability Inquiry; Other.\\
\addlinespace[2pt]
  & \dimname{query\_formality}
  & Register of the query text on a colloquial-to-elevated
    spectrum.
  & single, ord., $|\mathcal V|=4$
  & Unknown; Informal; Neutral; Formal.\\
\addlinespace[2pt]
  & \dimname{query\_structure}
  & Surface organization of the query: terse imperative, flowing
    prose, or explicit organizational markers (lists, headers).
  & single, nom., $|\mathcal V|=4$
  & Unknown; Direct; Verbose; Structured.\\
\addlinespace[2pt]
  & \dimname{query\_technical}
  & Sophistication of the language used in the query (an observable
    linguistic property, not an inference about the user).
  & single, ord., $|\mathcal V|=4$
  & Unknown; Non Technical; Intermediate; Expert.\\
\addlinespace[2pt]
  & \dimname{query\_engagement}
  & How the user positions themselves toward the agent: requesting
    a deliverable, asking for advice, delegating, or seeking to
    understand.
  & multi, nom., $|\mathcal V|=5$
  & Unknown; Directive; Consultative; Delegating; Inquisitive.\\
\addlinespace[2pt]
  & \dimname{query\_constraints}
  & How tightly the user constrains the response, measured by the
    count and explicit organization of stated requirements.
  & single, ord., $|\mathcal V|=5$
  & Unknown; Minimal; Moderate; Detailed; Comprehensive.\\
\midrule
\multirow{3}{=}{\textit{Tone \& Affect}}
  & \dimname{tone\_sentiment}
  & Primary affective polarity of the query.
  & single, ord., $|\mathcal V|=6$
  & Unknown; Strongly Negative; Negative; Neutral; Positive;
    Strongly Positive.\\
\addlinespace[2pt]
  & \dimname{tone\_emotion}
  & Emotional undertones inferable from the language; multiple
    emotions may co-occur.
  & multi, nom., $|\mathcal V|=9$
  & Unknown; Enthusiasm; Curiosity; Frustration; Impatience;
    Confusion; Uncertainty; Neutral; Other.\\
\addlinespace[2pt]
  & \dimname{tone\_urgency}
  & Time pressure signalled by the query, from no temporal context
    to explicit emergency markers.
  & single, ord., $|\mathcal V|=5$
  & Unknown; Low; Medium; High; Critical.\\
\midrule
\multirow{5}{=}{\textit{Output}}
  & \dimname{output\_type}
  & For document-producing intents, the kind of artifact requested
    (\dimval{Unknown} when no document is expected).
  & single, nom., $|\mathcal V|=11$
  & Unknown; Report; Proposal; Presentation; Spreadsheet; Plan;
    Documentation; Letter; Email; Tutorial; Other.\\
\addlinespace[2pt]
  & \dimname{output\_format}
  & Concrete file format for the requested artifact.
  & single, nom., $|\mathcal V|=6$
  & Unknown; Word; PowerPoint; Excel; PDF; Other.\\
\addlinespace[2pt]
  & \dimname{output\_length}
  & User-stated or implied expectation for the response/artifact
    length.
  & single, ord., $|\mathcal V|=5$
  & Unknown; Brief; Moderate; Detailed; Extensive.\\
\addlinespace[2pt]
  & \dimname{output\_components}
  & Specific components the user expects in the output (only
    meaningful when a document is being produced).
  & multi, nom., $|\mathcal V|=6$
  & Code; Examples; Sources; Visuals; Actionable; Other.\\
\addlinespace[2pt]
  & \dimname{output\_creativity}
  & Expected stylistic register of the output: factual, balanced
    (default for most document requests), or imaginative.
  & single, ord., $|\mathcal V|=4$
  & Unknown; Factual; Balanced; Creative.\\
\midrule
\multirow{3}{=}{\textit{Content}}
  & \dimname{domain\_names}
  & Subject-matter domain(s) of the query; multiple domains may
    apply.
  & multi, nom., $|\mathcal V|=21$
  & Technology Computing; Engineering; Natural Sciences;
    Mathematics; Medical Health; Psychology Cognitive; Business
    Economics; Law Legal; Education Pedagogy; Arts Humanities;
    Creative Arts; Social Sciences; Language Linguistics; Lifestyle
    Personal; Entertainment Media; Sports Recreation; Food
    Culinary; Travel Geography; Environment Sustainability;
    Spirituality Religion; Other.\\
\addlinespace[2pt]
  & \dimname{content\_specificity}
  & How concretely the query is anchored to real entities, numbers,
    and references.  Generic descriptors (``our company'', ``the
    team'') do not count as grounding.
  & single, ord., $|\mathcal V|=4$
  & Unknown; Abstract; Grounded; Highly Specific.\\
\addlinespace[2pt]
  & \dimname{content\_embedded}
  & Non-natural-language artifacts embedded inline in the query
    text.
  & multi, nom., $|\mathcal V|=5$
  & Code; Formula; Data; URL; Other.\\
\midrule
\multirow{4}{=}{\textit{Observable}}
  & \dimname{char\_count\_bucket}
  & Query length in characters, bucketed.  Measured directly from
    the query text.
  & single, ord., $|\mathcal V|=6$
  & Unknown; 1--50; 50--100; 100--200; 200--500; $\geq 500$.\\
\addlinespace[2pt]
  & \dimname{word\_count\_bucket}
  & Query length in words, bucketed.  Measured directly from the
    query text.
  & single, ord., $|\mathcal V|=6$
  & Unknown; 1--10; 10--20; 20--40; 40--100; $\geq 100$.\\
\addlinespace[2pt]
  & \dimname{query\_language}
  & Natural language of the query, detected from the text.
  & single, nom., $|\mathcal V|=11$
  & Unknown; English; Spanish; French; German; Italian; Portuguese;
    Japanese; Mandarin; Hindi; Other.\\
\addlinespace[2pt]
  & \dimname{grounding\_explicit}
  & Types of grounding artifacts (attached files, referenced
    entities) the user provides alongside the query.
  & multi, nom., $|\mathcal V|=20$
  & \texttt{pptx}; \texttt{docx}; \texttt{xlsx}; \texttt{pdf};
    \texttt{csv}; \texttt{md}; \texttt{txt}; \texttt{json};
    \texttt{html}; \texttt{aspx}; \texttt{png}; \texttt{image};
    \texttt{file}; \texttt{page}; \texttt{loop}; \texttt{meeting};
    \texttt{email}; \texttt{chat}; \texttt{people}; Other.\\
\end{longtable}
}

\section{Roundtrip Metric Formulae}\label{sec:appendix-metrics}

This appendix gives the full definitions of the three families of
quantities used in the roundtrip analysis (\S\ref{sec:eval-roundtrip},
\S\ref{sec:eval-refinement}): the per-dimension proxy distance, the
proportional-redistribution confusion matrix, and the size-adjusted
diagonal-concentration score.  All classifier confidence scores are
integers in $[1, S]$, with $S$ the schema's maximum confidence
(typically $S = 5$); the distinguished value \dimval{Unknown} always
carries score~$0$.

\subsection{Per-Dimension Distance}\label{sec:appendix-distance}

A proxy assigns each dimension either a single \texttt{[value, score]}
pair or, for multi-valued dimensions, a list of such pairs.

\paragraph{Single-valued dimensions.}
Let $(v_1, s_1)$ and $(v_2, s_2)$ be the original and reconstructed
pairs.

\noindent\textit{(i)~one side is \textup{Unknown}:}
$d = s_{\text{other}} / S$ (and $d = 0$ if both are Unknown).
\noindent\textit{(ii)~$v_1 = v_2$:}
$d = |s_1 - s_2| / S$ (pure confidence mismatch).
\noindent\textit{(iii)~$v_1 \neq v_2$, both real:}
\begin{equation}\label{eq:single-distance}
  d = \min\!\bigl(1,\; f_{\text{value}} \cdot f_{\text{score}}\bigr),
  \quad
  f_{\text{score}} = \frac{s_1 + s_2}{2S},
\end{equation}
where $f_{\text{value}} = 1$ for nominal dimensions and, for ordinal
dimensions of arity $n$,
\begin{equation}\label{eq:ordinal-mismatch}
  f_{\text{value}}
    = \frac{|\operatorname{rank}(v_1) - \operatorname{rank}(v_2)|}
           {(n - 1)/2}.
\end{equation}
Adjacent ordinal values therefore incur only a partial penalty, while
values more than half the range apart saturate to~1; high-confidence
mismatches are penalized more than low-confidence ones.

\paragraph{Multi-valued dimensions.}
Let $O$ and $R$ be the original and reconstructed value--score sets,
with score lookups $s^O_v, s^R_v$.  Partition the value union into
shared, original-only, and reconstructed-only labels.  If
$\text{shared} = \emptyset$ and at least one side is non-empty,
$d = 1$.  Otherwise,
\begin{equation}\label{eq:multi-distance}
  d = \min\!\left(1,\;
    \frac{
      \displaystyle\sum_{v\in\text{shared}} \tfrac{|s^O_v - s^R_v|}{S}
      + \sum_{v\in\text{orig-only}}  \tfrac{s^O_v}{S}
      + \sum_{v\in\text{recon-only}} \tfrac{s^R_v}{S}
    }{|\text{shared}|}
  \right).
\end{equation}
Dividing by the number of shared values lets correct matches buy
tolerance for additional mismatches.

\paragraph{Proxy-level aggregation.}
The scalar proxy distance $d \in [0,1]$ used in
Table~\ref{tab:roundtrip} is a weighted mean of the per-dimension
distances across all non-observable dimensions, using the same
\emph{raw} importance weights $w_i$ that enter the RA score
(\S\ref{sec:drift}), i.e.\ before the redundancy discount
$g_i$ of (\ref{eq:ra-score}) is applied.  Per-proxy aggregation has
no notion of cross-dimension overlap to discount, so only the
operator-supplied importance weights are used.

\subsection{Confusion-Matrix Construction}\label{sec:appendix-confusion}

For a single roundtrip line with original values
$\{(v_k, s_k)\}_{k=1}^{K}$ and reconstructed values
$\{(u_l, t_l)\}_{l=1}^{L}$, mass is added to the per-dimension
matrix $C \in \mathbb{R}_{\geq 0}^{|V|\times|V|}$ in four steps:

\begin{enumerate}
  \item \textbf{Score floor for Unknown.}  Replace each
    \dimval{Unknown}'s score by $\max(s, 1)$ so that it contributes
    visible mass; weights elsewhere are $w_k = s_k$.
  \item \textbf{Balance.}  Let
    $S_O = \sum_k w_k$ and $S_R = \sum_l w'_l$.  Append a synthetic
    \texttt{<blank>} pseudovalue with weight
    $|S_O - S_R|$ to the lighter side; the total mass to distribute
    is $M = 2\max(S_O, S_R)$.
  \item \textbf{Diagonal (matches).}  For each value $v$ present on
    both sides with weights $(w, w')$, add $2\min(w, w')$ to $C_{v,v}$
    and subtract $\min(w, w')$ from each side's residual.
  \item \textbf{Off-diagonal (proportional).}  Distribute the
    remaining mass $M_{\text{rem}}$ across all (original,
    reconstructed) pairs in proportion to the residual marginals:
    \begin{equation}\label{eq:offdiag}
      C_{v_k, u_l} \mathrel{+}= M_{\text{rem}}\,
        \frac{w_k}{\sum_{k'} w_{k'}}\,
        \frac{w'_l}{\sum_{l'} w'_{l'}}.
    \end{equation}
\end{enumerate}

\noindent
After accumulating across all $M\!\times\!N$ roundtrip samples, each
row is normalized so that $\sum_j C_{ij} = 1$.

\subsection{Diagonal Concentration}\label{sec:appendix-concentration}

Each confusion matrix is summarized by a scalar
\textbf{concentration around the main diagonal}.  We first compute a
raw score and then apply a size adjustment.

\paragraph{Raw score.}
For \emph{nominal} dimensions the raw score is the diagonal fraction,
\begin{equation}\label{eq:concentration-nominal}
  s_{\text{raw}} = \frac{\sum_i C_{ii}}{\sum_{i,j} C_{ij}}.
\end{equation}
For \emph{ordinal} dimensions of arity $n$ each cell receives a
proximity weight:
\begin{equation}\label{eq:weights}
  w_{ij} =
  \begin{cases}
    1 - \tfrac{|i-j|}{n-1}, & \text{both ordered values at ranks } i, j,\\
    1, & \text{both \dimval{Unknown}},\\
    0.5, & \text{exactly one side \dimval{Unknown}},\\
    0, & \text{either side is \texttt{<blank>}},
  \end{cases}
\end{equation}
yielding
\begin{equation}\label{eq:concentration-ordinal}
  s_{\text{raw}} =
    \frac{\sum_{i,j} C_{ij}\, w_{ij}}{\sum_{i,j} C_{ij}}.
\end{equation}
Blank entries (lost or hallucinated labels) carry the maximum
penalty; \dimval{Unknown} on only one side is treated as a half
mismatch, reflecting partial (not definitive) information.

\paragraph{Size adjustment.}
Larger matrices spread mass across more cells, making high raw scores
harder to achieve; we compensate with a size-dependent exponent,
\begin{equation}\label{eq:concentration-size-adjustment}
  s_{\text{adj}} = s_{\text{raw}}^{\,1/\sqrt{n}},
\end{equation}
which leaves $0$ and $1$ fixed and boosts intermediate scores in
proportion to $\sqrt{n}$.  At $n = 10$, a raw score of $0.60$ maps to
$0.81$.

\paragraph{Macro vs.\ micro averaging.}
The \emph{micro} variant applies
(\ref{eq:concentration-nominal})/(\ref{eq:concentration-ordinal}) to
the full mass-weighted matrix, so frequent values dominate.  The
\emph{macro} variant computes the per-row diagonal share for each
value and averages them with equal weight, exposing how rare values
behave.  All ``macro diagonal concentration'' figures in
\S\ref{sec:eval-refinement} use the macro variant.

\section{Unbiased Mutual Information for Discriminative Power}
\label{sec:appendix-ami}

The discriminative-power evaluation of \S\ref{sec:eval-dimpower}
requires safeguards against the upward bias of
mutual information on finite samples.
Disjoint training and validation folds already control CLL gain.
However, for ordinal dimensions, we
must additionally choose between the raw partition and a
PAV-merged coarsening on the \emph{same} training fold; this
nested comparison cannot be settled by held-out CLL alone, so we
use an Adjusted Mutual Information (AMI) tiebreaker.

\paragraph{Need for chance correction.}
Under $X \indep Y$, the plug-in estimator
$\widehat{\MI}(X;Y) = \sum_{ij}\hat p_{ij}\log(\hat
p_{ij}/(\hat p_{i\cdot}\hat p_{\cdot j}))$ is strictly positive in
expectation, with bias roughly $(R{-}1)(C{-}1)/(2N)$ that grows
with the partition cardinality $R\!\times\!C$.  For nested
candidates the data-processing inequality already pins the
direction (the finer partition wins in the population), so the
test is one-sided and any uncorrected positive bias
deterministically flips the verdict toward the finer partition.

\paragraph{Vinh--Epps--Bailey adjusted MI.}
We adopt the closed-form expectation under a fixed-margin
hypergeometric null derived in \cite{vinh2010ami}:
\begin{equation}\label{eq:vinh-emi}
\E[\MI \mid \mathbf{a}, \mathbf{b}, N]
= \sum_{i,j} \sum_{n_{ij}=\ell_{ij}}^{u_{ij}}
  \frac{n_{ij}}{N}\,\log\!\frac{N\,n_{ij}}{a_i b_j}\,
  \mathbb{P}[N_{ij}{=}n_{ij}\mid a_i, b_j, N],
\end{equation}
where $(\ell_{ij}, u_{ij}) = (\max(0, a_i{+}b_j{-}N), \min(a_i,
b_j))$ and $\mathbb{P}[\cdot]$ is the central hypergeometric mass.
The chance-corrected score is
\begin{equation}\label{eq:ami-max}
\AMI_{\max}(X;Y)
= \frac{\MI(X;Y) - \E[\MI]}{\max(H(X), H(Y)) - \E[\MI]},
\end{equation}
clipped to $[0,1]$.  We use the $\max$ normalizer because it is
the only one Vinh~et~al.\ prove respects the unit interval under
the hypergeometric null.  The inner sum is evaluated in log-space
via a precomputed $\log\Gamma$ table to avoid factorial overflow.

\paragraph{Kish correction for fractional weights.}
Equation~(\ref{eq:vinh-emi}) requires \emph{integer} margins and an
integer total $N$.  Our contingencies are weighted by
classifier-confidence scores $w_e\in(0,1]$, so $N=\sum_e w_e$ is
generally non-integer.  We replace it with Kish's effective sample
size \cite{kish1965survey},
\begin{equation}\label{eq:kish}
N_{\mathrm{eff}}
= \frac{\bigl(\sum_e w_e\bigr)^{2}}{\sum_e w_e^{2}},
\end{equation}
rounded to the nearest integer.  Intuitively, $N_{\mathrm{eff}}$
shrinks when a few events carry most of the weight; the null
distribution should be wider in that regime, and the AMI
adjustment correspondingly more aggressive.  Weighted margins are
scaled by $N_{\mathrm{eff}}/N$ and rounded with the largest-remainder
method to obtain integer pseudo-margins
$(\mathbf{a}^{*}, \mathbf{b}^{*})$ summing to $N_{\mathrm{eff}}$.
These are substituted for $(\mathbf{a}, \mathbf{b}, N)$ throughout
the right-hand side of (\ref{eq:vinh-emi}).  Only $\E[\MI]$ is
computed this way; the observed $\MI(X;Y)$ and the entropies in
(\ref{eq:ami-max}) come from the original weighted contingency.
The substitution is the conventional adjustment when applying a
counting null to a weighted sample and errs conservatively
(slightly inflating $\E[\MI]$ and so deflating $\AMI$).

\paragraph{Limited application for AMI.}
We use AMI only as the train-fold tiebreaker between the raw and
PAV-refined partitions for ordinal dimensions, where a like-for-like
comparison that controls for partition cardinality is essential.
However, the operational question ``how much would knowing $X$
help me predict $Y$ on a new interaction?'' is answered directly
by held-out CLL gain, which is what
\S\ref{sec:eval-dimpower} reports.